\documentclass[aps,reprint,floatfix,superscriptaddress,texcount]{revtex4-1}

\usepackage[usenames,dvipsnames]{xcolor}
\usepackage{amsmath,amssymb,amsthm,bbold}

\definecolor{icolor1}{HTML}{0000FF}
\definecolor{icolor2}{HTML}{FFA400}
\definecolor{icolor3}{HTML}{008B00}
\definecolor{icolor4}{HTML}{FF34A7}
\definecolor{icolor5}{HTML}{632280}
\definecolor{icolor6}{HTML}{5E3100}

\newcommand{\Dthreed}{\mathrm{D}_\mathrm{3d}}

\definecolor{ourBlue}{HTML}{0000b8}
\definecolor{ourOrange}{HTML}{FFA400}
\definecolor{ourGreen}{HTML}{008B00}
\definecolor{ourMagenta}{HTML}{FF34A7}
\definecolor{ourPurple}{HTML}{632280}
\definecolor{ourBrown}{HTML}{5E3100}
\definecolor{ourRed}{HTML}{b90000}

\usepackage{graphicx}
\usepackage[separate-uncertainty=true, multi-part-units=single]{siunitx}
\usepackage{braket}
\newcommand*\autoref[1]{Figure \ref{#1}}
\usepackage{hyperref}
\hypersetup{
    colorlinks,
    linkcolor={black},
    citecolor={blue},
    urlcolor={blue}
}

\begin{document}

\title{Initialization and Readout of Nuclear Spins via negatively charged Silicon-Vacancy Center in Diamond}

\author{Mathias~H.~Metsch}
\author{Katharina~Senkalla}
\affiliation{Institute for Quantum Optics, Ulm University, D-89081 Germany}
\author{Benedikt~Tratzmiller}
\affiliation{Institute for Theoretical Physics, Ulm University, D-89081 Germany}
\author{Jochen~Scheuer}
\affiliation{Institute for Quantum Optics, Ulm University, D-89081 Germany}
\author{Michael~Kern}
\affiliation{Institute for Quantum Optics, Ulm University, D-89081 Germany}
\author{Jocelyn~Achard}
\affiliation{Universit\'e Paris 13, Sorbonne Paris Cit\'e, LSPM-CNRS (UPR3407), 93430 Villetaneuse, France}
\author{Alexandre Tallaire}
\affiliation{Universit\'e Paris 13, Sorbonne Paris Cit\'e, LSPM-CNRS (UPR3407), 93430 Villetaneuse, France}
\affiliation{Institut de Recherche de Chimie Paris, Chimie ParisTech, CNRS, PSL Research University, 75005 Paris, France}
\author{Martin~B.~Plenio}
\affiliation{Institute for Theoretical Physics, Ulm University, D-89081 Germany}
\affiliation{Center for Integrated Quantum Science and Technology (IQ$^\text{{st}}$), Ulm University, D-89081 Germany}
\author{Petr~Siyushev}
\email{petr.siyushev@uni-ulm.de}
\affiliation{Institute for Quantum Optics, Ulm University, D-89081 Germany}
\author{Fedor~Jelezko}
\affiliation{Institute for Quantum Optics, Ulm University, D-89081 Germany}
\affiliation{Center for Integrated Quantum Science and Technology (IQ$^\text{{st}}$), Ulm University, D-89081 Germany}

\begin{abstract}
In this work, we demonstrate initialization and readout of nuclear spins via a negatively charged silicon-vacancy (SiV) electron spin qubit. 
%
%
Under Hartmann-Hahn conditions the electron spin polarization is coherently transferred to the nuclear spin. 
The readout of the nuclear polarization is observed via the fluorescence of the SiV.
We also show that the coherence time of the nuclear spin (6 ms) is limited by the electron spin-lattice relaxation due to the hyperfine coupling to the electron spin.
This work paves the way towards realization of building blocks of quantum hardware with an efficient spin-photon interface based on the SiV color center coupled to a long lasting nuclear memory. 
\end{abstract}


\maketitle



\begin{figure}
	\includegraphics[width=0.9\columnwidth]{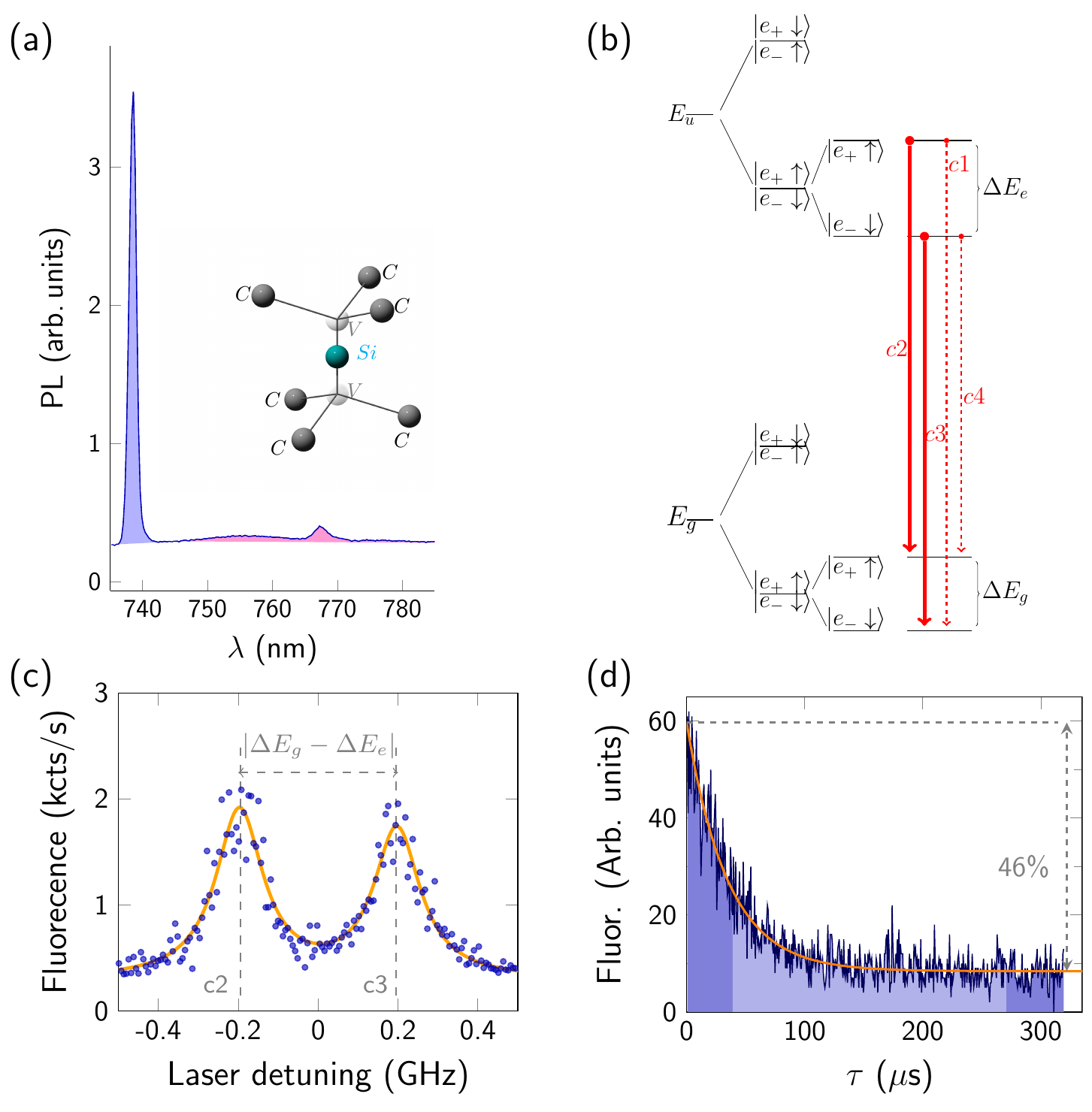}
	\caption{
	(a) Photoluminescence spectrum of a SiV center at $2\,\mathrm{K}$.  
	The blue-shaded region highlights the prominence of the zero-phonon line which yields up for $70\%$ of the total fluorescence. 
	The pink region indicates weak phonon sideband.
	Inset: The physical structure of the defect composed of an interstitial Si atom in between of two vacancies.
	(b) Energy level scheme of a SiV center. 
	Only the lowest Zeeman sublevels in the ground and excited states are shown. 
	Spin conserving optical transitions between Zeeman sublevels are denoted by the solid red arrows, whereas red dashed arrows point spin flipping transition.
	(c) PLE spectrum obtained at $188.7\,\mathrm{mT}$. 
	The peaks correspond to the transitions indicated in (b) as $c2$ and $c3$. 
	The observed separation of the transitions is $380\,\mathrm{MHz}$, which corresponds to the difference in Zeeman splitting of the ground and excited states. 
	(d) Polarization of the electron spin of the defect by a laser pulse resonant to the transition $c3$. 
	Fit to the data yields a polarization of the spin of $\approx 92\%$.
	Dark blue-shaded areas around leading the edge peak and falling edge of the steady state fluorescence are integrated and used in all pulsed experiments as measurement and normalization signals, respectively.
		}
	\label{fig:figure1}
\end{figure}

\begin{figure}
	\includegraphics[width=0.9\columnwidth]{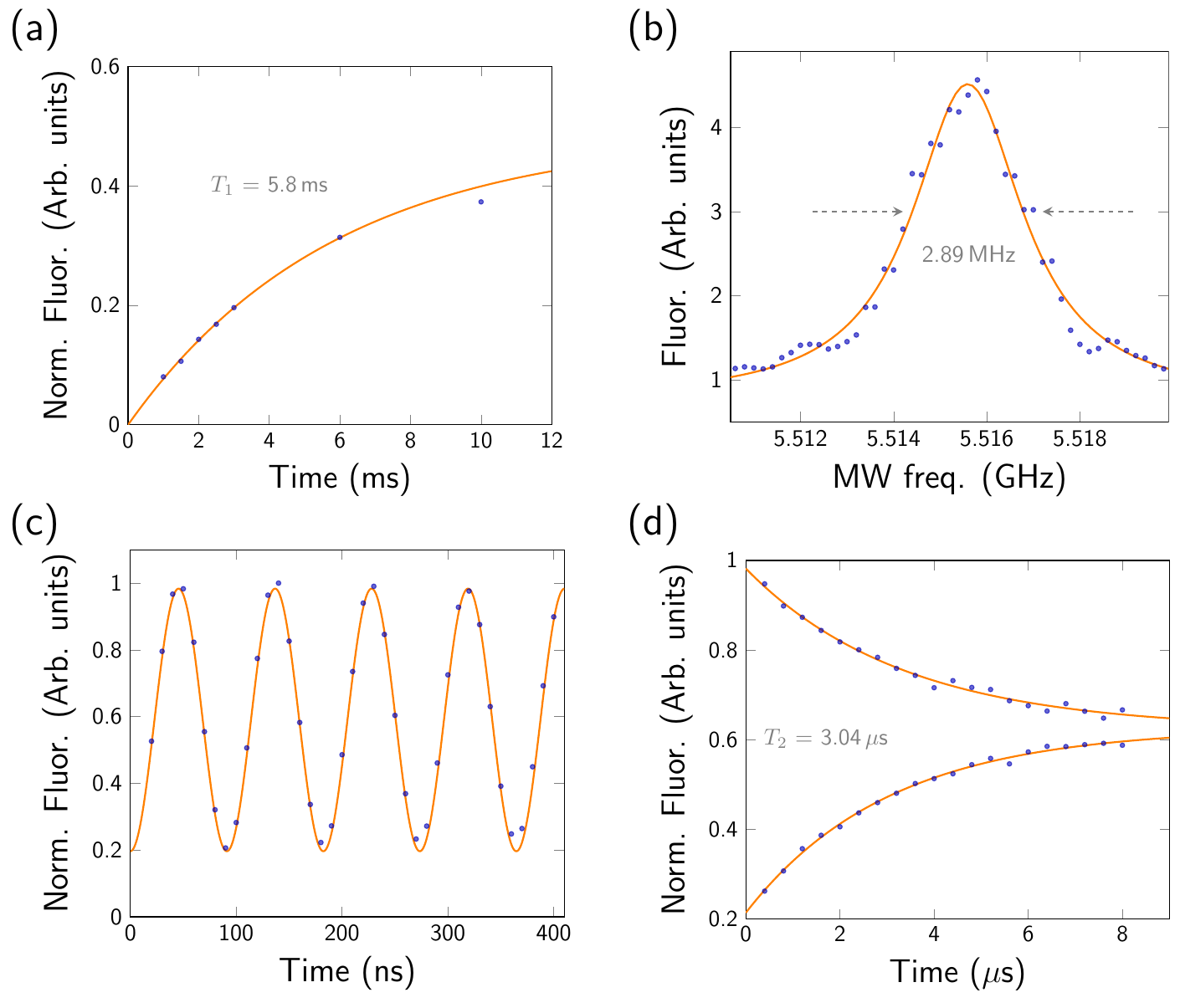}
	\caption{
	(a) Electron spin lifetime of the center used in presented experiments is $\approx5.8\,\mathrm{ms}$.
	(b) ODMR of the electron spin revealed a width of $2.89\,\mathrm{MHz}$.
	(c) $10\,\mathrm{MHz}$ Rabi oscillation of the electron spin. 
	(d) Hahn-echo decay measurements yields a $\mathrm{T}_2\approx 3\,\mu\mathrm{s}$. 
	It is more than three orders of magnitude shorter than $\mathrm{T}_1$ time.
		}
	\label{fig:figure2}
\end{figure}

\begin{figure}
	\includegraphics[width=0.9\columnwidth]{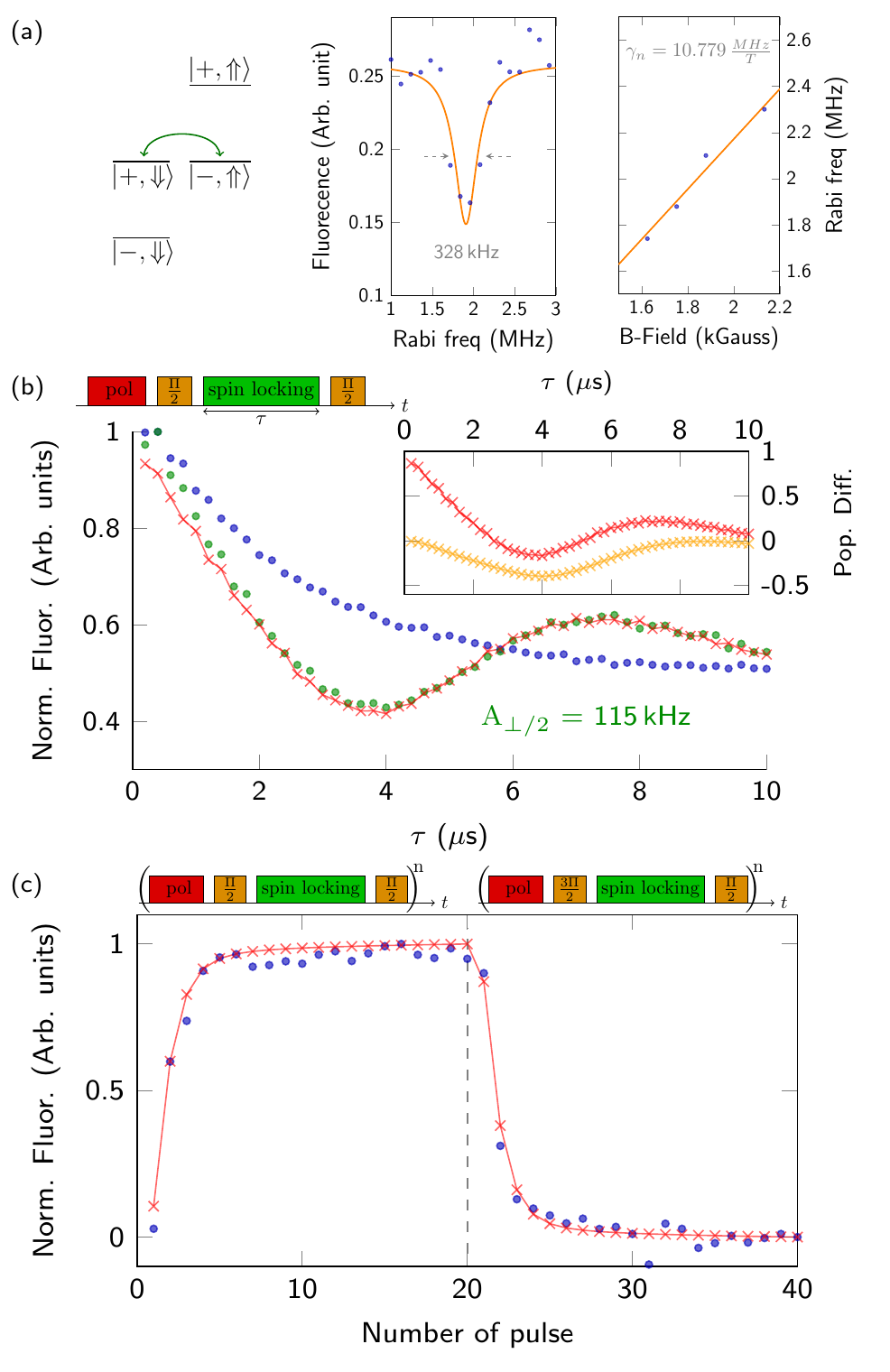}
	\caption{
	(a) Schematic representation of the joint electron-nuclear spin dynamics explaining Hartmann-Hahn double resonance, spin-locking and dynamic nuclear polarization (see text for more details).
	Green arrow denotes the coherent oscillation between $\ket{+,\Downarrow}$ and $\ket{-,\Uparrow}$ states.
      The observed Hartmann-Hahn double resonance exhibits a width of $328\,\mathrm{kHz}$ and its resonance frequency has a linear dependence on the external magnetic field.
	The $\gamma_N$, obtained from the linear fit, is $10.7\,\mathrm{MHz/T}$ which allows us to assign coupling to $^{13}\mathrm{C}$ nuclei spins $(\gamma _{C13} = 10.705\,\mathrm{MHz/T})$.
	(b) Two spin-locking decays obtained at Rabi frequency not matching (blue) and matching (green) Hartmann-Hahn condition.
	 Spin-locking performed at resonance shows evidence of coupling to a single nuclear spin (green), it is in well agreement with the simulation (red).
The measurement was performed in an alternating fashion.
	 Inset: Simulated population of the electron spin (red) and coupled $^{13}\mathrm{C}$ spin (yellow) showing a transfer of polarization to the nuclear spin.
	(c) Repetitive transfer of the electron spin polarization to the surrounding nuclear spins using NOVEL technique.
		}
	\label{fig:figure3}
\end{figure}

\begin{figure}
	\includegraphics[width=0.9\columnwidth]{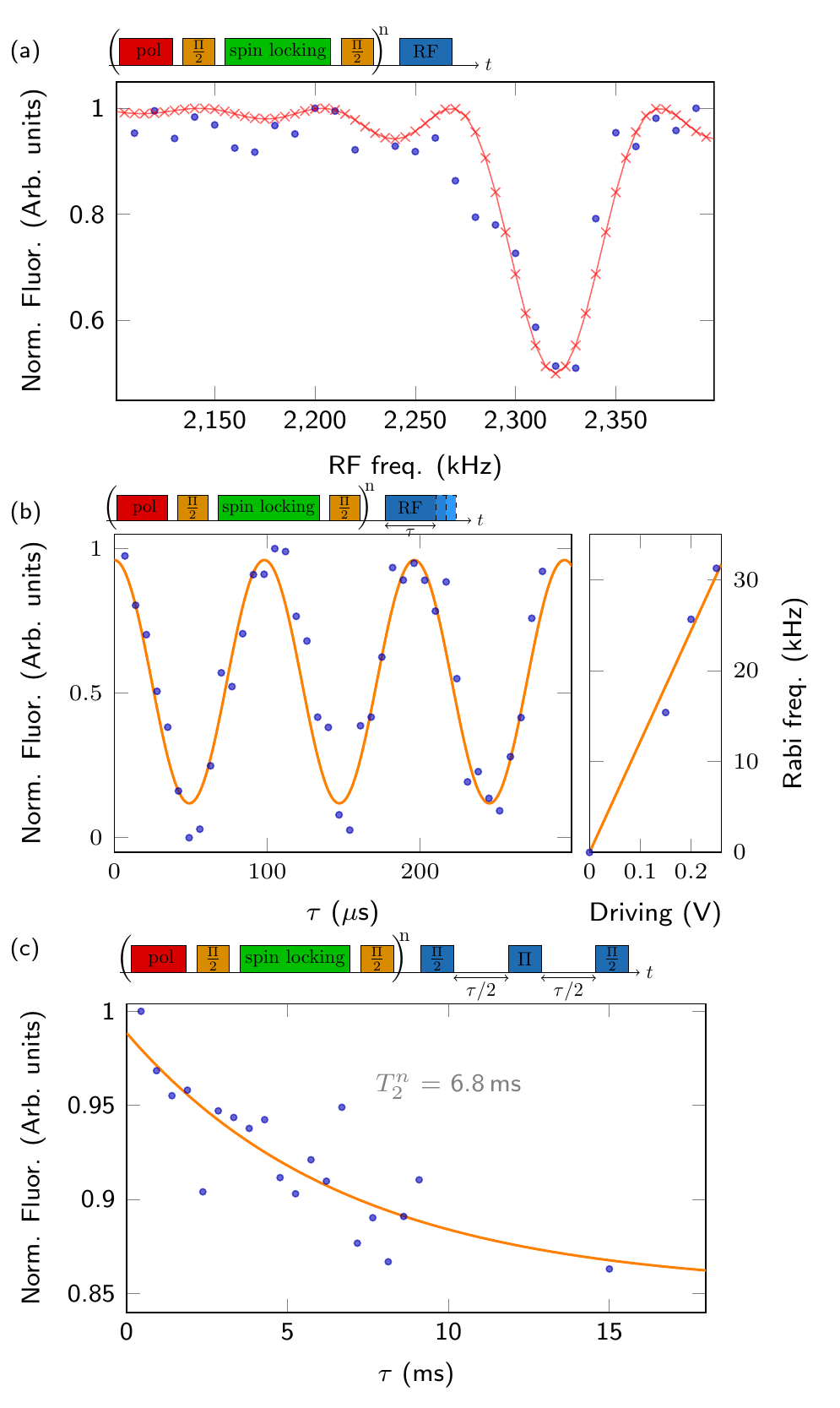}
	\caption{	
	(a) Measurement of the NMR signal of the individual nuclear spin (blue) with simulation results (red).
	The observed resonance at $2.32\,\mathrm{MHz}$ is shifted compared to the Larmor frequency of $^{13}\mathrm{C}$ obtained by Hartmann-Hahn double resonance due to dipolar interaction between electron and nearby nuclear spin ($\mathrm{A}_{\parallel}/2 = 360\,\mathrm{kHz}$).
	%
	(b) Left: Nuclear Rabi flops $(10.1\,\mathrm{kHz})$. 
	Right: Linear amplitude dependency of Rabi frequency.
	(c) The measurement of the coherence time of the nuclear spin via spin-echo decay yields $\mathrm{T}^{n}_{2}=6.8\,\mathrm{ms}$, comparable to the $\mathrm{T}_{1}$ of the electron spin.
	%
	%
}
	\label{fig:figure4}
\end{figure}

Recent advances with color centers in diamond based on IV-group  elements \citep{Hepp_PRL2014,Palyanov_SR2015,Iwasaki_PRL2017,Tchernij_ACSP2017} hold promise to provide an efficient interface between photons and spin qubits.
These color centers possess a high Debye-Waller factor (larger than $0.5$) \citep{Vlasov_AM2009,Neu_NJP2011}, which implies a high flux of coherent photons, and furthermore an exceptional spectral stability owing to the inversion symmetry of the defects \citep{Sipahigil_PRL2014,Siyushev_PRB2017}.
Both of these properties are crucial to realize long distant entanglement based on light-matter interface
\citep{AtomStateTeleportation1999,Moehring_N2007,santori2002indistinguishable} which forms an essential building block
for scalable quantum processors and quantum repeaters.
However, these color centers are not free of constraints.
A main drawback is the limited coherence time of the electron spin which is induced by a fast phonon mediated relaxation process between the orbital branches of the ground state \citep{Jahnke_NJP2015}.
Different attempts to overcome this problem were recently demonstrated, including the application of high strain \citep{Sohn_NC2018} or freezing of the specimen to millikelvin temperatures \citep{Sukachev_PRL2017}.
Of these two methods, the former leads to symmetry distortion that affects the optical properties, while the latter requires dilution refrigerators, which are expensive and offer only limited cooling capability.
Another route to beat this limitation is to use a defect of this family only as a spin-photon interface and readout gate, while storing the information on a long living nuclear memory.
However, to realize such a hybrid approach several problems have to be tackled.
Among them are the initialization of the nuclear memory and its readout.
The simple polarization technique utilizing level anticrossing and optical pumping commonly used for NV center coupled to $^{13}\mathrm{C}$ or ST1 centers \citep{Jacques_PRL2009,Lee_NN2013} is not applicable to the systems of SiV family.\\
In this letter, we report deterministic polarization of a small nuclear spin ensemble via a dynamic nuclear polarization protocol.
By measuring the Larmor precession in different magnetic fields we identify these nuclei as $^{13}\mathrm{C}$.
From this ensemble, we choose one nuclear spin with coupling strength in the order of few hundred $\mathrm{kHz}$.
This $^{13}\mathrm{C}$ spin is used to demonstrate nuclear magnetic resonance and Rabi oscillations via nuclear spin polarization readout protocol.\\
%
%
For the experiment, a $\langle 111 \rangle$-oriented diamond sample containing ingrown SiV center was chosen and placed in a low temperature confocal microscope \citep{BINDER201785} operating at $\sim2\,\mathrm{K}$.
The SiV centers were resonantly excited by a tunable single frequency Ti:sapphire laser.
The detection of the emitted fluorescence was restricted to the phonon sideband, ranging from approximately $ 750\,\mathrm{nm} $ to $810\,\mathrm{nm}$ (\autoref{fig:figure1}~a) to exclude the reflection of the excitation laser.
To increase photon collection efficiency, a solid immersion lens (SIL) was fabricated on the sample surface.
The magnetic field, produced by a superconducting coil, could be applied along the $\langle 111 \rangle$-direction of diamond, parallel to the excitation laser beam.
To provide microwave and radiofrequency fields a $20\,\mu\mathrm{m}$ thick copper wire was spanned across the sample in the vicinity of the SIL.\\
The SiV center consists of an interstitial Si atom aligned along $\langle 111 \rangle$ direction in between of two vacancies (inset of \autoref{fig:figure1}~a), resulting in $\Dthreed$ point group symmetry that defines the electronic structure of the defect~\citep{Hepp_PRL2014}.
The ground $^{2}\mathrm{E}_{g}$ and excited $^{2}\mathrm{E}_{u}$ state of the color center are orbital doublets, which are split by spin-orbit interaction (\autoref{fig:figure1}~b).
Due to the negative charge of the defect, the total electron spin is one-half, resulting in spin degeneracy of each orbital state.
This degeneracy is lifted in external magnetic fields and the Zeeman sublevels can be treated as a spin qubit \citep{rogers2014all-optical}.
These sublevels in the ground and excited states are linked via spin conserving transitions.
Due to the small difference of g-factors in the ground and excited state, transitions between states with opposite spin projections become resolvable and can be separately addressed. 
An example of the photoluminescence excitation (PLE) spectrum at the magnetic field of $188.7\,\mathrm{mT}$ is depicted in \autoref{fig:figure1}~c.
When the magnetic field is aligned along the principal axis of the defect, the spin projection is a good quantum number, which leads to long cycling transitions.
However, small misalignment of the magnetic field results in a mixture of different spin components and thus the probability of relaxation into another Zeeman sublevel increases.
This effect was used to achieve spin initialization by optical pumping (\autoref{fig:figure1}~d).
To observe these dynamics, resonant laser pulses of $300\,\mu\mathrm{s}$ were applied and time-resolved fluorescence was recorded.
The difference between the fluorescent level at the beginning of the pulse and at its end indicates 92~\% spin polarization.
The initialization fidelity can be further improved by a larger separation of the optical transitions in increased magnetic fields or by the use of a spin flipping transition for the initialization \citep{rogers2014all-optical}, so that the laser operating on one transition does not induce repumping from the state with opposite spin projection. \\
The spin-lattice relaxation time $(\mathrm{T}_1)$ was evaluated by applying an additional laser pulse after variable interpulse interval.
This measurement yields a $\mathrm{T}_1$ time exceeding $5\,\mathrm{ms}$ (\autoref{fig:figure2}~a).
When the polarized center is subjected to a microwave field resonant to the splitting of the Zeeman sublevels, population between them is exchanged and thus fluorescence on the optically pumped transition is increased.
A typical optically detected magnetic resonance (ODMR) with a linewidth of $2.89\,\mathrm{MHz}$ is shown in~\autoref{fig:figure2}~b.
%
%
Observing the fluorescence depending on the length of a resonant MW-pulse leads to Rabi oscillation, which provides the timing for $\pi$-pulse on the electron spin~(\autoref{fig:figure2}~c).
The coherence time was measured by Hahn echo decay (\autoref{fig:figure2}~d) and found to be $\mathrm{T}_{2}\approx 3\,\mu\mathrm{s}$.
The coherence time is longer than the $\mathrm{T}^{*}_{2}$ time, which can be estimated from the ODMR linewidth to be of the order of $100\,\mathrm{ns}$.
%
%
Although, the ODMR line for this particular center does not exhibit any obvious coupling to the surrounding nuclear spins, it can still be used as a probe for its nuclear spin environment.
To identify the Larmor frequency ($\omega_L$) of the nuclear spins in the applied field we utilized the well established technique known as Hartmann-Hahn double resonance~\citep{Hartmann1962}.
The idea of this method is to enable cross relaxation between two spin systems with large energy mismatch by driving one system with a Rabi frequency which is in resonance with the energy of another spin system, as depicted in \autoref{fig:figure3}~a.
Experimentally, after initialization of the SiV electron spin in the $|\uparrow\rangle$ state, it is rotated by a microwave $\pi/2$-pulse into superposition state $|+\rangle = 1/\sqrt{2}(|\uparrow\rangle+|\downarrow\rangle)$.
Followed by another long microwave pulse, phase shifted by $90^{\circ}$ in respect to the previous, such that the spin is locked.\\
When the MW amplitude of the the spin-locking ($90^{\circ}$-phase shifted) pulse corresponds to a Rabi frequency ($\Omega$) which is matching the condition $\Omega = \gamma_N|\mathbf{B}|\equiv\omega_L$, where $\gamma_N$ is the nuclear gyromagnetic ratio and $\mathbf{B}$ is the applied magnetic field, the Hartmann-Hahn condition is fulfilled and cross relaxation occurs.
This resonance condition can be probed via measuring the remaining population of the electron spin in the coherent superposition state $|+\rangle$, which is transferred to populations by the second $\pi/2$-pulse followed by optical readout.
Varying the amplitude of this spin-locking pulse, the nuclear precession frequency is observed (\autoref{fig:figure3}~a). 
As expected, the position of the resonance changes linearly with external magnetic field $\mathbf{B}$ and thus exclude artifact signals (\autoref{fig:figure3}~a). 
From the slope of the curve, we extract the nuclear gyromagnetic ratio $\gamma_N=10.7\,\mathrm{MHz/T}$. 
This matches well the gyromagentic ratio of ${}^{13}\mathrm{C}$ \citep{bernstein2004handbook}.\\
%
%
We should stress that Hartmann-Hahn resonance condition for the interacting $1/2$-spin systems is only defined by the Larmor frequency and does not permit to distinguish different nuclei coupled to the probe spin.
%
%
This is in contrast to NV center (spin 1) coupled to an $^{13}\mathrm{C}$ ensemble~\citep{London_PRL2013}, where the resonance condition is shifted by a value proportional to the hyperfine coupling $A_{\parallel}$.\\
However, we can reveal the presence of a stronger coupled nuclear spin using the spin-locking technique. 
The sequence used for this purpose is similarly to the described above but the duration of the spin-locking pulse is varied.
If the amplitude of this spin-locking pulse is chosen such that the Rabi frequency is not in resonance with the precession frequency of the nuclear spins no cross relaxation occurs.
The electron spin is effectively decoupled from the environmental magnetic noise so that the coherence decays with its characteristic time $\mathrm{T}_{1\rho}$~\citep{Redfield_PR1955}.
As described above, the dynamics changes if the Rabi frequency of the spin-locking pulse coincides with the nuclear Larmor frequency: Cross relaxation leads to decoherence due to interaction with multiple spins. 
However, if the SiV electron spin is coupled to just one nucleus, the coupled system evolves coherently exhibiting oscillatory behavior.
The non-resonant and resonant cases of the spin-locking decay curves are depicted in \autoref{fig:figure3}~b. 
At resonance oscillations are clearly visible, which suggest that the coupling to one individual $^{13}\mathrm{C}$ is dominant.
%
To describe the dynamics of the nuclear ($\vec{I}$) and the electron spin ($\vec{S}$) during the spin-locking pulse, we used the following Hamiltonian
	\begin{equation}
		H = \Omega S_x + \omega_L I_z + A_\perp S_z I_x + A_\parallel S_z I_z \, ,
		\label{eq:eq1}
	\end{equation}
where $\Omega$ is the spin-locking frequency, $A_\perp$ and $A_\parallel$ are the perpendicular and parallel components of the hyperfine tensor (see \autoref{fig:figure3} and \autoref{fig:figure4}) .
The evolution of the joint system is governed by the Lindblad equation with SiV $\mathrm{T}_2$ time included in the relaxation term	
	\begin{equation}
	\dot{\rho}  = -i \left[H, \rho\right] + \frac{1}{2 T_2} \left((\sigma_z\otimes\mathbb{1}) \rho (\sigma_z\otimes\mathbb{1}) - \rho \right).
	\label{eq:eq2}
	\end{equation}
The simulated process assuming a single coupled  $^{13}\mathrm{C}$ appear to be in good agreement with the measured data, as depicted in \autoref{fig:figure3}~b.
The inset of \autoref{fig:figure3}~b shows the separately simulated evolution of the electron (red curve) and nuclear (yellow curve) spins.\\
Moreover, the application of this sequence at resonant conditions leads to the polarization of the nuclear spins.
%
The spin-locking measurements (\autoref{fig:figure3}~b) revealed $\mathrm{A}_{\perp}$ for this particular ${}^{13}\mathrm{C}$ to be $(2\pi)\times 230\,\mathrm{kHz}$.
Thus, optimal polarization transfer to the nuclear spin would occur at $4.35\,\mu \mathrm{s}$.
However, the joint electron-nuclear state decays on the $\mathrm{T}_2$ timescale reducing nuclear spin polarization.
For the given system, optimal polarization transfer is achieved with a spin-locking time of $\approx 4\,\mu\mathrm{s}$ (inset of \autoref{fig:figure3}~b).
\\
Multiple repetitions of this sequence with an appropriate length of the spin-locking pulse results in accumulation of polarization of nuclear spin bath as shown in  \autoref{fig:figure3}~c. 
Using the model described above, we estimate the degree of polarization of the strongest coupled ${}^{13}\mathrm{C}$ to be in excess of $60\% $.
The degree of polarization achieved is limited by the losses in a single polarization step due to the electron spin decay.
This technique is called nuclear spin orientation via electron spin-locking (NOVEL) and is well known in the field of nuclear magnetic resonance~\citep{Henstra_JMR1988}.
To record the previously described Hartmann-Hahn and spin-locking curves we had to avoid nuclear spin polarization, which would suppress contrast of the signals. 
This was achieved by reversing the polarization after each step, so that net nuclear polarization remains zero.\\
In order to distinguish the individual nuclear spin from the bath revealed by the spin-locking technique, we performed a nuclear magnetic resonance (NMR) measurement with readout via dynamical nuclear polarization  \citep{2017robustnucleispinpolarization,Unden2018}. 
After polarization of the nuclei, a radiofrequency (RF) pulse is applied followed by the polarization sequence that measures changes in nuclear polarization produced by the RF pulse. 
The RF frequency was swept around the Larmor precession frequency ($1.96\,\mathrm{MHz}$ at $188.7\,\mathrm{mT}$ magnetic field) in the range equal to the ODMR linewidth, since the coupling strength is smaller than $1/T^{*}_{2}$ of the electron spin.
We observe a change of the fluorescence level at a frequency of $2.32\,\mathrm{MHz}$, indicating the position of NMR signal of the single $^{13}\mathrm{C}$ nuclear spin (\autoref{fig:figure4}~a).
This frequency is shifted due to the dipolar coupling of the electron and the nuclear spins (as described in \autoref{eq:eq1}), which enables the addressing of a single nuclear spin although it is not directly resolved and therefore can be considered as weakly coupled.\\ 
To demonstrate the coherent manipulation of the single nuclear spin, we performed Rabi oscillation under RF driving.
The RF pulse of varying length at the resonance frequency of $2.32\,\mathrm{MHz}$ was applied in between two polarizing sequences as shown in \autoref{fig:figure4}~b.
The Rabi frequency increase linearly with the square root of RF power (\autoref{fig:figure4}~b).\\
With a defined RF $\pi$-pulse, we evaluated the coherence time of the $^{13}\mathrm{C}$.
The echo decay was measured by varying free evolution time separating RF $\pi$-pulse from $\pi/2$-pulses (\autoref{fig:figure4}~c). It revealed a decay of $\mathrm{T}^{n}_2=6.8\,\mathrm{ms}$, which is similar to the measured electron $\mathrm{T}_1$ time of SiV.
To counteract the decay of the electron polarization, the electron spin state is reinitialized just prior to every RF pulse.
This is important for sequences with long waiting times, since for the two different electron spin orientations the RF transition frequency would be found either at $\omega_{L } + A_{\parallel}/2$ or $\omega_{L} - A_{\parallel}/2$.
It should be noticed that the ability to manipulate relatively weak coupled nuclear spin is highly beneficial for implementation of quantum computational protocols, since operation on auxiliary electron spin, e.g. its (re)initialization, does not lead to fast decoherence of the nuclear storage qubit.
For the given coupling, $\approx16\%$ loss of the nuclear polarization was observed due to the re-initializing laser pulses.
As before, initialization and readout were implemented via NOVEL sequence.
Considering the coupling strength for this particular $^{13}\mathrm{C}$ to the SiV, a flip of the electron spin is the dominant source of decoherence of the nuclear spin.
The electron spin-lattice relaxation time can be enhanced by improved alignment of the magnetic field along the symmetry axis of SiV center, which would lead to a prolonged nuclear $\mathrm{T}^{n}_2$ time.
Furthermore, it might be possible to apply dynamic decoupling to the nuclear spin to suppress decoherence.
\\
In conclusion, we demonstrated  initialization and readout of a nuclear spin by electron spin ancilla.
This work establishes a hybrid approach for realization of scalable quantum register or quantum networking that combines one of the best-known solid state spin-photon interface with long living nuclear memory.
Although the electron spin coherence time remains relatively short for information swap between electron and nuclear spins, other defects of the SiV family, such as GeV or SnV, are expected to have better coherent properties due to larger orbital splitting of the ground state.
A longer electron spin coherence would also enable addressing of weaker coupled $^{13}\mathrm{C}$ which would reduce the decoherence of the nuclear spin due to a flip of the electron spin.
This would enable the use of such hybrid system at liquid $^{4}\mathrm{He}$ temperatures without needing expensive dilution refrigerators and simplify experimental technique. 


%
%
%



\section*{Acknowledgements}
This work was supported by the Landesstiftung BW, the DFG, the ERC Synergy grant BioQ, the EU project ASTERIQS and HYPERDIAMOND, the BMBF via NanoSpin, DiaPol and Q.Link.X and  the Paris \^{\i}le-de-France Region in the framework of DIM SIRTEQ.

\bibliography{siv_c13_bib}

\end{document}